\documentclass[prb,aps,superscriptaddress,preprint]{revtex4-2}

\usepackage{graphicx}
\usepackage{amsmath}
\usepackage[english]{babel}
\usepackage{units}
\usepackage[colorlinks=true,urlcolor=blue,linkcolor=blue,citecolor=blue]{hyperref}
\usepackage{mathptmx}
\usepackage{xcolor}

\begin{document} 

\title{Non-local electrical detection of spin-polarized surface currents in the \\ 3D topological insulator BiSbTeSe$_{2}$}

\author{Shaham Jafarpisheh}
\affiliation{2nd Institute of Physics and JARA-FIT, RWTH Aachen University, 52074 Aachen, Germany}
\affiliation{Peter Gr\"unberg Institute (PGI-9), Forschungszentrum J\"ulich, 52425 J\"ulich, Germany}

\author{Frank Volmer}
\affiliation{2nd Institute of Physics and JARA-FIT, RWTH Aachen University, 52074 Aachen, Germany}

\author{Zhiwei Wang}
\affiliation{Physics Institute II, University Cologne, 50937 Cologne, Germany}
\affiliation{Key Laboratory of Advanced Optoelectronic Quantum Architecture and Measurement, School of Physics, Beijing Institute of Technology, Beijing 100081, P. R. China}

\author{B\'arbara Canto}
\affiliation{AMO GmbH,Gesellschaft f\"ur Angewandte Mikro- und Optoelektronik, 52074 Aachen, Germany}

\author{Yoichi Ando}
\affiliation{Physics Institute II, University Cologne, 50937 Cologne, Germany}

\author{Christoph Stampfer}
\affiliation{2nd Institute of Physics and JARA-FIT, RWTH Aachen University, 52074 Aachen, Germany}
\affiliation{Peter Gr\"unberg Institute (PGI-9), Forschungszentrum J\"ulich, 52425 J\"ulich, Germany}

\author{Bernd Beschoten}
\thanks{e-mail: bernd.beschoten@physik.rwth-aachen.de}
\affiliation{2nd Institute of Physics and JARA-FIT, RWTH Aachen University, 52074 Aachen, Germany}
 
\begin{abstract}
	The spin-polarized surface states in topological insulators offer unique transport characteristics that make them distinguishable from trivial conductors. Here, we detect the impact of these surface states in the topological insulator BiSbTeSe$_2$ by electrical means using a non-local transport configuration with ferromagnetic Co/Al$_2$O$_3$ electrodes. We show that the non-local measurement allows to probe the surface currents flowing along the whole surface, i.e.~from the top along the side to the bottom surface and back to the top surface along the opposite side. Increasing the temperature increases the interaction between bulk and surface states, which shortens this non-local current path along the surface and hence leads to a complete disappearance of the non-local signal at around 20~K. Interestingly, we observe that the ratio between spin signal to background signal is much larger in the non-local geometry compared to the local one. Given that the observed ratio in the non-local geometry aligns well with expectations for spin-polarized surface states, our findings suggest that an as-yet unresolved mechanism diminishes the spin signal in the local geometry.  
\end{abstract}

\maketitle

Topological surface states (TSS) and two-dimensional electron gases (2DEG) can coexist at the surface of three-dimensional topological insulators (TIs)~\cite{Bianchi2010,Michiardi2022,Interplay-between-Topological-States-and-Rashba-States}. Whereas the TSS exhibit spin-momentum locking, the 2DEG is characterized by bands with pronounced Rashba spin splitting~\cite{PhysRevLett.107.096802}. Under an electrical bias voltage, the spin-textures of these surface states lead to a current-induced spin polarization (CISP), where the polarization is perpendicular to the momentum direction, however with different signs between the TSS and Rashba-split states (RSS)~\cite{Hasan2010Nov,Hong2012Aug,Ando2013Sep,PhysRevB.103.035412,PhysRevApplied.20.024065}. Therefore, both TSS and RSS offer many opportunities for spintronics and magnetoelectronics~\cite{Topological-spintronics-and-magnetoelectronics,Opportunities-in-topological-insulator-devices,PhysRevApplied.20.024065,Li2022}. For such purposes, spin-sensitive ferromagnetic electrodes can be used to indirectly detect the spin-polarized surface currents, as these currents create a difference in the electrochemical potentials for spins parallel and antiparallel to the magnetization direction of the electrodes~\cite{Tian2015Sep,ncomms13518,Li2019May}. Therefore, reversing either the magnetization direction of the ferromagnetic electrodes by an external magnetic field or reversing the flow direction of the current lead to a sign reversal in the magnetic hysteresis loop of the detected spin signal. This behavior has indeed been observed in several studies before~\cite{PhysRevB.103.035412,PhysRevApplied.20.024065,Li2014Feb,Dankert2015Dec,Tang2014Sep,Ando2014Nov,HWANG2019917,Vaklinova2018Feb,Dankert2018Mar,Tian2015Sep,Lee2015Oct,Yang2016Aug}.

\begin{figure*}[tb]
	\includegraphics[width=\linewidth]{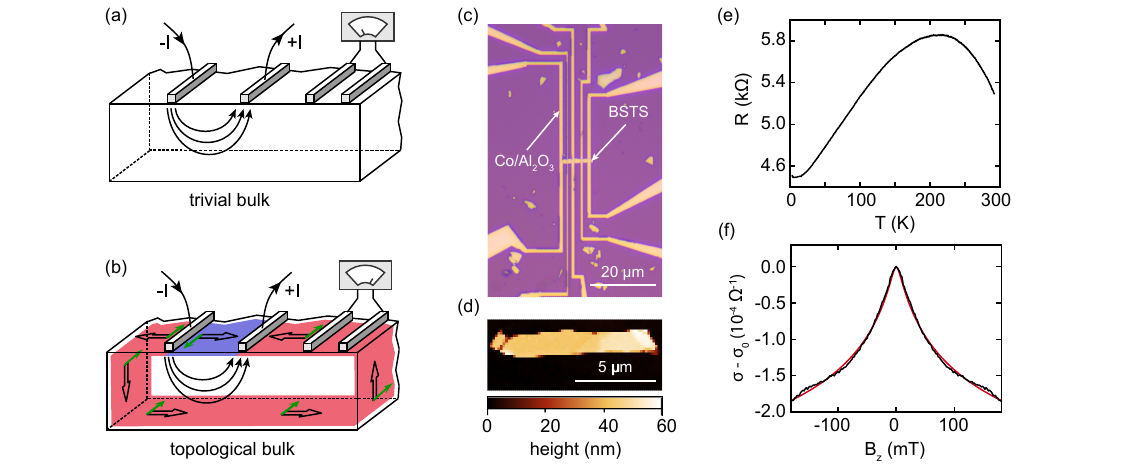}
	\caption{\label{fig:figure1}
		(a) Distribution of the current inside the bulk of a material with trivial band structure (black arrows). Only a negligible amount of current due to the effect of current spreading is expected under the non-local detection electrodes. (b) Distribution of the current inside a 3D topological insulator. Depicted are the potentially present bulk channel (black arrows), the local surface channel directly between source and drain (blue), and a second surface channel (red), which connects source and drain in a non-local manner. 
        The spin polarization directions are indicated by green arrows and correspond to the case where spin-momentum–locked transport carried by the topological surface states (TSS) dominates over transport via the Rashba-split states (RSS). (c) Optical image of the 40~nm thick exfoliated BSTS flake contacted with Co/Al$_2$O$_3$ electrodes. (d) AFM scan of the flake shown in panel (c) before electrode fabrication. (e) Four-probe resistance as a function of temperature. (f) Weak antilocalization measurement at a base temperature of $T$~=~3~K as a function of an out-of-plane magnetic field $B_z$. The red line is a fit using the Hikami-Larkin-Nagaoka model.}
\end{figure*}

However, so far the amplitudes of reported spin signals in local measurement geometries are surprisingly small for allegedly fully polarized surface currents and are additionally superimposed by magnetization independent background signals that are two to three orders of magnitude larger than the spin signal~\cite{Tian2015Sep,PhysRevB.103.035412,Li2014Feb,Dankert2015Dec,Tang2014Sep,Ando2014Nov,HWANG2019917}. One possible explanation for such diminished spin signals are invasive contacts that might disturb the spin polarized surface states~\cite{Culcer2020,Majumder2017,Walsh2017,PhysRevB.90.085115,Longo2025}. Another explanation might be possible contributions from transport over bulk and/or defect states in the local transport geometry (black arrows in Figs.~\ref{fig:figure1}a and \ref{fig:figure1}b)~\cite{PhysRevLett.121.176801,doping-bulk,Shirodkar2022,PhysRevB.108.115401,Sasmal_2021,Jafarpisheh21}. Charge transport over such states, especially in the presence of magnetic fringe fields, can also show similar magnetic-field dependent transport signatures compared to the signatures attributed to the TSS or RSS~\cite{Kuntsevich2018May,deVries2015Nov,PhysRevB.103.035412}. This can be a problem in local measurements of the surface states in-between source and drain contacts (transport over these surface states is depicted by the blue area in Fig.~\ref{fig:figure1}(b)). However, the topologically protected surface states as well as the Rashba-split states are expected to fully enclose the 3D bulk connecting all facets of a strong TI~\cite{Fu2007Mar}. If surface states on the one hand and bulk and defect states on the other hand are sufficiently decoupled, a third transport channel therefore opens from source to drain, carrying the current non-locally along the surface (see red current path in Fig.~\ref{fig:figure1}(b)) 
~\cite{Wolgast2013Nov,Kim2013Nov,Shekhar2014Oct,Lee2014Mar}. A non-local, spin-sensitive voltage measurement of this second surface transport channel thus offers in principle a way to separate the contributions from surface states on the one hand and bulk and/or defect states on the other hand.

In this work, we investigate both local (L) and non-local (NL) electrical transport measurements of the surface states in exfoliated BiSbTeSe$_{2}$ (BSTS) flakes~\cite{Ren2011Oct}. And indeed, we observe distinctive different spin transport properties in these two configurations. 
On the one hand, the ratio between the spin signal and the background signal is nearly two orders of magnitude higher in the non-local geometry compared to the local geometry, the latter being consistent with values reported in the literature.
This hints to the fact that so far an unknown phenomena diminishes the spin signal in the local geometry. On the other hand, we observe significantly different temperature dependencies between local and non-local spin signals. Thermally activated hopping transport is likely the reason for the different temperature dependencies, as the non-local current path is shortened by this temperature activated channel.

The BSTS flakes were prepared by mechanical exfoliation onto Si$^{++}$/SiO$_2$(300~nm) substrates. Tunnel-barriers were fabricated by evaporation of an aluminium (Al) seed layer ($\approx$~1~nm) over the whole flake which got naturally oxidized, followed by thermal atomic layer deposition (ALD) with tri-methylaluminum (TMA) and water as precursors for an additional $\approx$~1~nm Al$_2$O$_3$. Here, the ALD process was used to close pinholes in the naturally oxidized Al$_2$O$_3$ layer. In the next step, e-beam lithography and e-beam evaporation was used to fabricate the cobalt (Co) electrodes. An optical image of the device together with an atomic force microscope (AFM) image of the BSTS-flake after exfoliation are shown in Figs.~\ref{fig:figure1}(c) and \ref{fig:figure1}(d), respectively. The thickness of the flake is $\approx$~40~nm.

The temperature dependent resistance of the device, measured in a four-probe configuration, shows a semiconducting behavior at high temperatures and a transition to a metallic regime at $T$~$\approx$~200~K (see Fig.~\ref{fig:figure1}(e)). This transition corresponds to the depletion of impurity and defect bands within the bulk band gap~\cite{Ren2010Dec}. The metallic behavior, on the other hand, indicates surface transport that persists down to low temperatures. As it is depicted in Fig.~\ref{fig:figure1}(f), a low-field magneto-resistance measurement at $T$~=~3~K shows the cusp-like feature resulting from weak anti-localization (WAL)~\cite{Chiu2013Jan,Lu2011Sep,Lu2011Aug,Kim2011Aug,Kim2013Jun}. The existence of WAL in topological insulators is a hallmark of the strong spin-orbit coupling. 
We use the Hikami-Larkin-Nagaoka model~\cite{Hikami1980Feb,Ockelmann2015Aug} for fitting the WAL curve (see red trace in Fig.~\ref{fig:figure1}(f)) and extract a phase coherence length of around 240~nm. 

\begin{figure*}[tb]
	\includegraphics[width=\linewidth]{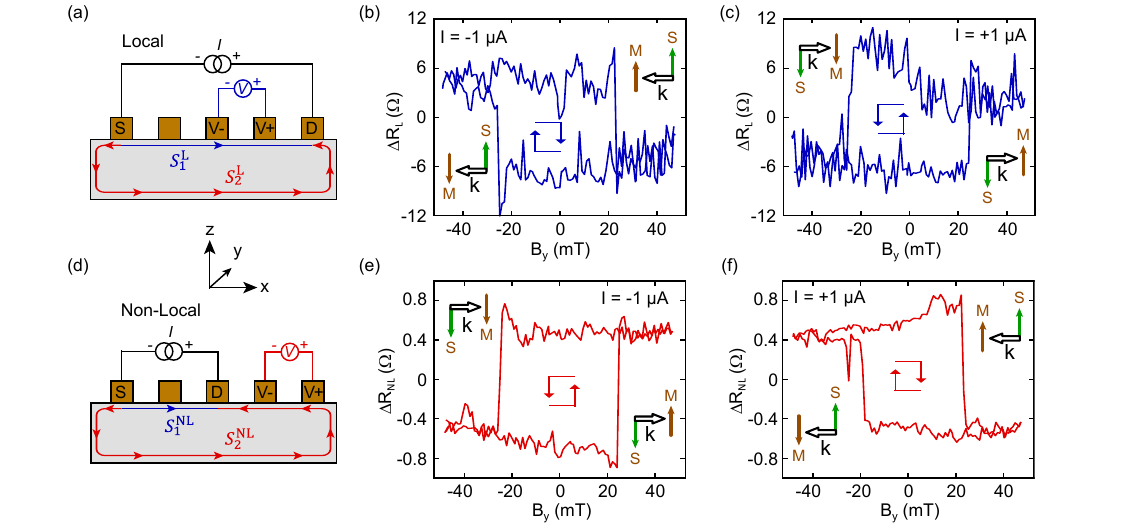}
	\caption{\label{fig:figure2}
		Schematic illustration of (a) local and (d) non-local measurement configurations. Potentiometric detection of the spin-polarization in the topological surface states using (b,c) local and (e,f) non-local geometries at $T$~=~3~K. The flow direction of electrons ($\vec{k}$) and their spin-polarization direction ($\vec{S}$) assuming the spin-texture of the topological surface states are shown by black and green arrows, respectively. In-plane magnetization direction of the electrodes ($\vec{M}$) is indicated using the brown arrows. All measurements show a hysteretic switching of the spin signal when reversing the magnetization direction of the ferromagnetic detectors by an in-plane external magnetic field $B_y$.}
\end{figure*}

In order to probe the spin polarized surface states, we now focus on transport measurements in both local and non-local detection geometries (see Figs.~\ref{fig:figure2}(a) and \ref{fig:figure2}(d), respectively), using spin sensitive ferromagnetic electrodes to allow for potentiometric detection~\cite{Hong2012Aug}. For this we apply in-plane magnetic fields along the $y$-direction to reverse the magnetization directions ($\vec{M}$) of the Co electrodes. We observe well-defined hysteresis loops in both local and non-local configurations which are expected when the contacts are magnetized either parallel or antiparallel to the spin-polarization ($\vec{S}$) of the surface charge carriers (Figs.~\ref{fig:figure2}(b,c) and~\ref{fig:figure2}(e,f)). These measurements were conducted by applying a DC current of $\pm$1~$\mu$A between source (S) and drain (D) electrodes. After the subtraction of a background signal, which will be discussed further below, we normalize the measured voltage with the applied current between source and drain to acquire the local and non-local resistances depicted in Figs.~\ref{fig:figure2}(b,c) and~\ref{fig:figure2}(e,f). In case of small currents as used in our measurements, it was shown that the spin-related voltage signal scales linearly with the applied current, before it saturates in some studies towards very high currents, possibly due to Joule heating or voltage-dependent population of non-surface states~\cite{PhysRevB.103.035412,PhysRevApplied.20.024065,Tian2015Sep,Li2014Feb,Dankert2015Dec,Dankert2018Mar,Ando2014Nov,HWANG2019917}. Within this linear range, calculating the local and non-local resistances is therefore an easy way to compare results from different studies, which normally report on local spin signals in the range of a few m$\Omega$ to 10~$\Omega$~\cite{PhysRevB.103.035412,PhysRevApplied.20.024065,Li2014Feb,Vaklinova2018Feb,Dankert2015Dec,Dankert2018Mar,Tian2015Sep,Lee2015Oct,Tang2014Sep,Ando2014Nov,Yang2016Aug,HWANG2019917}. The quite large spin signals of around 10~$\Omega$ in the local transport geometry in our study (see Figs.~\ref{fig:figure2}(b) and \ref{fig:figure2}(c)) therefore belong to the highest spin signals reported in literature so far.

Important contributors to this large spin signal are the optimized tunnel barriers that show contact-resistance–area products of around 20-30~k$\Omega\mu m^2$. This is in the same range in which previous studies both on topological insulators~\cite{Vaklinova2018Feb} and graphene~\cite{PhysRevB.90.165403} have demonstrated increased spin signals due to increased spin injection/detection efficiencies and reduction of contact-induced spin relaxation processes. Furthermore, we note that in contrast to most other studies all of our contacts are ferromagnetic, including the source and drain contacts. This might be important because the TSS is not expected to be 100\% spin-polarized, despite its spiral nature. This is due to spin-orbit interactions and a possible penetration of the TSS into subsurface layers, where fluctuating electric fields can result in complex spin textures~\cite{PhysRevLett.105.266806,PhysRevLett.110.216401,ARPES-review,arpes-polarization,PhysRevB.86.235106}. Furthermore, it is the topic of ongoing research how electrical contacts to the TI affect both topological surface states and Rashba-split states~\cite{Culcer2020,Majumder2017,Walsh2017,PhysRevB.90.085115,Longo2025}. Overall, it is very likely that both spin up and down transport channels have to be considered even for the TSS~\cite{Li2019May}. Accordingly, there is the possibility that in our study the use ferromagnetic source and drain electrodes might boost the spin polarization within the surface states and therefore the spin signal.

Consistent to surface state transport, we observe that reversing the electrons' momentum direction ($\vec{k}$) leads to a reversal of the spin-polarization direction of the surface carriers and therefore the reversal of the respective hysteresis curves in both the L and the NL configurations (compare left and right panels in Figs.~\ref{fig:figure2}(b,c) and~\ref{fig:figure2}(e,f). We note that we cannot draw any definitive conclusions as to whether the spin signal is due to transport via TSS or RSS. In many studies the sign of the hysteresis loop is used for this distinction~\cite{PhysRevB.103.035412,PhysRevApplied.20.024065,Li2014Feb,Dankert2018Mar,Tian2015Sep,Ando2014Nov,Yang2016Aug}. However, as it was pointed out in other studies, the sign can also change depending on the resistance and characteristics of the electrical contacts~\cite{Li2019May,Vaklinova2018Feb}, making such an assignment ambiguous. This is aggravated in our case by the fact that not a single ferromagnetic contact is used with otherwise non-magnetic reference electrodes. If only ferromagnetic electrodes are used, the sign of the spin signal can in principle also change depending on the specific spin injection/detection efficiencies of the individual contacts~\cite{Li2019May}. We discuss both the experimental similarities and dissimilarities between TSS and RSS and the resulting problems in assigning the observed spin signal to one of these states in more detail in the Supporting Information.

Next we focus on the results of the non-local measurements in Figs.~\ref{fig:figure2}(e) and \ref{fig:figure2}(f). The amplitude of the spin signal of 1~$\Omega$ is around one order of magnitude smaller than in the local geometry, but still in the higher range of reported values in literature. Interestingly, we observe also a reversal of the hysteresis when comparing local and non-local measurements for the same direction of the current between source and drain electrodes (compare respective upper and lower panels in Figs.~\ref{fig:figure2}(b,c) and~\ref{fig:figure2}(e,f). Under the assumption of similar spin detection efficiencies between the contacts, this sign reversal can be readily explained by a spin polarized current through the bottom surface (see $S_2^\mathrm{NL}$ and red arrows in Fig.~\ref{fig:figure2}(d)). As we kept the polarity of the voltage amplifier the same in both measurements geometries (first the negative input V$_-$ then the positive input V$_+$ along the x-axis, compare Figs.~\ref{fig:figure2}(a) and (d)), a current over $S_2^\mathrm{NL}$ will have a reversed current direction in the non-local geometry (it flows from V$_+$ towards V$_-$) compared to the current over $S_1^\mathrm{L}$ in the local geometry (there it flows from V$_-$ towards V$_+$). This results in a sign reversal and shows that the non-local measurement allows to probe the surface currents flowing along the whole surface, i.e.~from the top along the side to the bottom surface and back to the top surface along the opposite side. 

\begin{figure}[tb]
	\includegraphics[width=\linewidth]{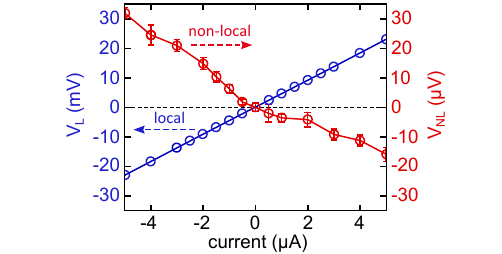}
	\caption{\label{fig:figure3}
	Non-magnetization-dependent background signals in both local (blue) and non-local (red) configurations as a function of the applied current between source and drain contacts.}
\end{figure}

\begin{figure*}[tb]
	\includegraphics[width=\linewidth]{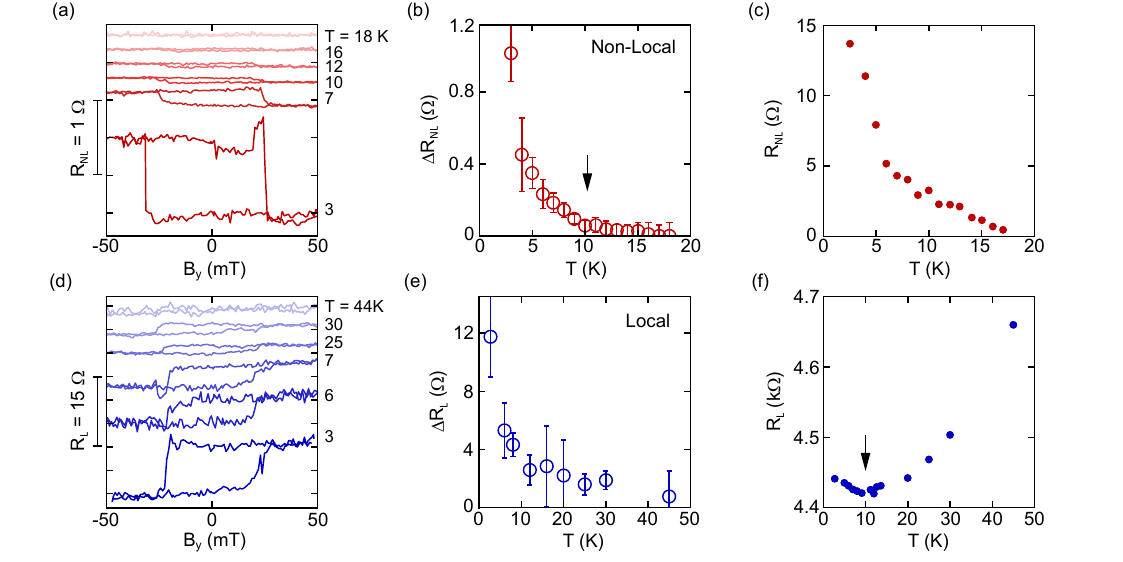}
	\caption{\label{fig:figure4}
    Temperature dependence of the (a) non-local and (d) local spin signals depicted as waterfall diagrams. Amplitudes of the spin signals in the (b) non-local and (e) local geometries as a function of temperature. The background signals on which the spin signal is superimposed in (a) and (d) are depicted in (c) for the non-local and in (f) for the local geometry by extracting the voltage values at B~=~50~mT from (a) and (d) and normalizing these values by the applied current between the source and drain electrodes. The black arrows indicate the temperature at which the local voltage in panel (f) shows a minimum and the non-local spin signal in panel (b) disappears.}
\end{figure*}

However, there is one striking difference between the local and non-local spin signals with respect to their magnetization independent background signals. Studies that conduct spin-sensitive measurements on TIs in the local detection geometry show background signals that are normally two to three orders of magnitude larger than the amplitude of the hysteresis signal~\cite{Tian2015Sep,PhysRevB.103.035412,Li2014Feb,Dankert2015Dec,Tang2014Sep,Ando2014Nov,HWANG2019917}. These background signals are normally subtracted from the spin-dependent data, as we have done in Fig.~\ref{fig:figure2}. The subtracted voltage signals are depicted in Fig.~\ref{fig:figure3} as a function of the applied current for both the local (blue) and non-local (red) geometry. The linear slope in the local geometry corresponds to a resistance of around 4.45~k$\Omega$, which is in accordance to the temperature dependent data in Fig.~\ref{fig:figure1}(e). Therefore, the amplitudes of the local spin signals in Figs.~\ref{fig:figure2}(b) and \ref{fig:figure2}(c) are around a factor of 450 smaller than this background signal, which is in good agreement to literature values. Instead, the slope in the non-local background signal in Fig.~\ref{fig:figure3}, whose sign reversal compared to the local configuration is indicative for a current flow in the -$x$ direction, corresponds to roughly 10~$\Omega$. This background signal is only one order of magnitude larger than the non-local spin signals in Figs.~\ref{fig:figure2}(e) and \ref{fig:figure2}(f). To the best of our knowledge, this is the largest spin signal to background ratio in spin-sensitive electrical measurements on TIs reported in literature so far.

It should be noted that the ratio between spin signal to background signal in the non-local geometry is not unreasonably large, but rather that the corresponding ratios in the local geometry are surprisingly low. If we assume that the spin polarization in the topological surface states is somewhere between 50\% to 100\%~\cite{PhysRevLett.105.266806,PhysRevLett.110.216401,ARPES-review,arpes-polarization,PhysRevB.86.235106} and if we assume typical spin detection efficiencies of ferromagnetic contacts with tunnel barriers of around 10\% to 30\%~\cite{Ahn2020,HAN2012369,PhysRevLett.105.167202,PhysRevB.80.214427}, one would expect that the non-magnetization-dependent background would indeed be only one order of magnitude larger than the spin signal. Why the ratio in the local geometry is significantly lower not only in our study but also in other studies in literature (the aforementioned two to three orders of magnitude ~\cite{Tian2015Sep,PhysRevB.103.035412,Li2014Feb,Dankert2015Dec,Tang2014Sep,Ando2014Nov,HWANG2019917}) should be the focus of further studies.

Next, we discuss the temperature dependence of both the spin and background signals in both measurement geometries. Figs.~\ref{fig:figure4}(a) and \ref{fig:figure4}(d) are waterfall diagrams depicting the evolution of the hysteresis curves for both local and non-local detection geometries with increasing temperature. The extracted amplitudes of the temperature dependent hysteresis curves are shown in Figs.~\ref{fig:figure4}(b) and \ref{fig:figure4}(e), respectively. The magnetic field independent background signals are plotted in Figs.~\ref{fig:figure4}(c) for the non-local and \ref{fig:figure4}(f) for the local geometry. There are two main observations in the temperature dependent measurements: First, the amplitude of the hysteresis in both configurations gets rapidly diminished as the temperature is increased. Interestingly, the non-local spin signal already vanishes around $T$~$\approx$~16~K (Fig.~\ref{fig:figure4}(b)) while the local spin signal lasts up to $T$~$\approx$~45~K (Fig.~\ref{fig:figure4}(e)). Assuming that both signals originate from the same TSS or RSS states, we must link the different temperature dependencies to the different current paths ($S_1^\mathrm{L}$ and $S_2^\mathrm{NL}$ in Figs.~\ref{fig:figure2}(a) and \ref{fig:figure2}(d), respectively).

In this respect, it is important that there is a distinct difference between the evolution of the background signals and the corresponding spin signals. In the non-local geometry both quantities ($\Delta R_\mathrm{NL}$ and $R_\mathrm{NL}$) show a monotonic decrease and disappear at similar temperatures (Figs.~\ref{fig:figure4}(b) and \ref{fig:figure4}(c)). Instead, the background signal in the local geometry ($R_\mathrm{L}$ in Fig.~\ref{fig:figure4}(f)) first slightly decreases from 3~K to 10~K before it increases towards higher temperatures. Interestingly, the minimum of the local resistance in Fig.~\ref{fig:figure4}(f) corresponds quite well with the temperature at which the non-local spin signal in Fig.~\ref{fig:figure4}(b) becomes very small (indicated by black arrows). As noted before, the local background signals extracted from the magnetic field dependent spin measurements (Fig.~\ref{fig:figure4}(f)) follow the temperature dependent resistance measurement of Fig.~\ref{fig:figure1}(e). The latter reveals a transition from activated to metallic transport below $T \approx$~200~K, which indicates the depletion of bulk states. However, as it was shown previously, variable range hopping (VRH) in the bulk may still contribute to transport at low temperatures~\cite{Skinner2012Oct,Ren2011Oct}. In particular, in BSTS compounds the compensation of acceptors and donors, which can explain the low bulk carrier concentration, results in random local potential fluctuations and therefore the formation of charged puddles~\cite{Skinner2012Oct}. Charge carriers can tunnel between neighboring puddles and can therefore form a parallel conduction channel which couples to the surface states and gives rise to additional scattering of surface carriers. This mechanism can limit the spin-polarized transport on the surface and therefore can lead to a suppression of the non-local signal~\cite{Velkov2018Oct}. We therefore argue that increasing the temperature toward 10~K activates hopping transport in the bulk, thereby shortening the non-local current path $S_2^\mathrm{NL}$ in Fig.~\ref{fig:figure2}(d), which ultimately leads to a suppression of the non-local spin signal at temperatures well below those relevant for the local transport geometry.

In conclusion, we have probed the spin-polarized surface states in BiSbTeSe$_{2}$ by both local and non-local spin-sensitive measurements. Our data suggest that the non-local measurement allows to probe spin-polarized surface currents flowing along the whole surface, i.e.~from the top along the side to the bottom surface and back to the top surface along the opposite side. This transport channel is suppressed at temperatures higher than 10~K most likely due to variable range hopping over bulk states, which shortens the non-local current path along the surface. Overall, our study demonstrates that at very low temperatures the non-local surface currents of a TI in contact to an insulator (SiO$_2$ in our case) should to be taken into account in any spin transport measurement. Finally, we observe that the ratio between spin signal to background signal is almost two orders of magnitude higher in the non-local geometry compared to the local one. In fact, the ratio in the non-local geometry is in quite good agreement to typical spin detection efficiencies of ferromagnetic contacts with tunnel barriers and the expected spin polarization of the surface states. Accordingly, our study hints to the fact that so far a not well understood phenomena is diminishing the spin signal in the local geometry, which exact origin should be the subject of further studies.

\begin{acknowledgments}
We gratefully acknowledge helpful discussions with M. Morgenstern. This project has received funding from the European Union's Horizon 2020 research and innovation programme under grant agreement No 785219 and No 881603 (Graphene Flagship) and No 796388 (ECOMAT), the Virtual Institute for Topological Insulators (J\"ulich-Aachen-W\"urzburg-Shanghai), the Deutsche Forschungsgemeinschaft (DFG, German Research Foundation) under Germany's Excellence Strategy - Cluster of Excellence Matter and Light for Quantum Computing (ML4Q) EXC 2004/1 - 390534769 (Gef\"ordert durch die Deutsche Forschungsgemeinschaft im Rahmen der Exzellenzstrategie des Bundes und der L\"ander - Exzellenzcluster Materie und Licht f\"ur Quanteninformation (ML4Q) EXC 2004/1 - 390534769), through DFG/SPP 1666 (BE 2441/8-2), and by the Helmholtz Nano Facility~\cite{Albrecht2017}. The work at Cologne was furthermore funded by the DFG under project number 277146847 - CRC 1238 (Subprojects A04) - and project number 398945897.
\end{acknowledgments}

\end{document}


\title{Supporting Information:\\ Non-local electrical detection of spin-polarized surface currents in the \\ 3D topological insulator BiSbTeSe$_{2}$}

\author{Shaham Jafarpisheh}
\affiliation{2nd Institute of Physics and JARA-FIT, RWTH Aachen University, 52074 Aachen, Germany}
\affiliation{Peter Gr\"unberg Institute (PGI-9), Forschungszentrum J\"ulich, 52425 J\"ulich, Germany}

\author{Frank Volmer}
\affiliation{2nd Institute of Physics and JARA-FIT, RWTH Aachen University, 52074 Aachen, Germany}

\author{Zhiwei Wang}
\affiliation{Physics Institute II, University Cologne, 50937 Cologne, Germany}
\affiliation{Key Laboratory of Advanced Optoelectronic Quantum Architecture and Measurement, School of Physics, Beijing Institute of Technology, Beijing 100081, P. R. China}

\author{B\'arbara Canto}
\affiliation{AMO GmbH,Gesellschaft f\"ur Angewandte Mikro- und Optoelektronik, 52074 Aachen, Germany}

\author{Yoichi Ando}
\affiliation{Physics Institute II, University Cologne, 50937 Cologne, Germany}

\author{Christoph Stampfer}
\affiliation{2nd Institute of Physics and JARA-FIT, RWTH Aachen University, 52074 Aachen, Germany}
\affiliation{Peter Gr\"unberg Institute (PGI-9), Forschungszentrum J\"ulich, 52425 J\"ulich, Germany}

\author{Bernd Beschoten}
\affiliation{2nd Institute of Physics and JARA-FIT, RWTH Aachen University, 52074 Aachen, Germany}

\maketitle

\begin{figure}[tb]
	\includegraphics[width=\linewidth]{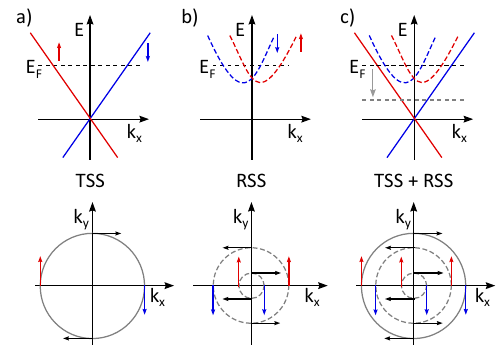}
	\caption{Simplified spin-polarized band structures around the $\Gamma$-point (electron energy E as a function of in-plane momentum k$_\text{x}$, upper panel) and the corresponding Fermi surface (lower pannel) of (a) topological surface states (TSS), (b) the Rashba-split states (RSS), and (c) the combination of both. Dashed black lines indicate the Fermi level E$_\text{F}$.}
    \label{SFigure1}
\end{figure}

In this Supporting Information we discuss the similarities and dissimilarities between  topological surface states (TSS) and Rashba-split states (RSS) with a particular focus on the experimental challenges in distinguishing between them. Figure~\ref{SFigure1} depicts simplified spin-polarized band structures and Fermi surfaces for both TSS and RSS (for a more detailed description of both types of states see references \cite{Hasan2010Nov,Rashba-Reference,Soumyanarayanan2016}).

Due to these spin-textures, an electrical bias voltage applied to the topological insulator creates a current-induced spin polarization (CISP), where the spin orientation is locked  perpendicular to the momentum direction. However, there are two important differences between the CISP created by TSS and RSS: the sign and the magnitude of the resulting  spin  polarization~\cite{Hasan2010Nov,Hong2012Aug,Ando2013Sep,PhysRevB.103.035412,PhysRevApplied.20.024065,PhysRevLett.107.096802}. As discussed in the main manuscript, early studies tried to distinguish the states by analyzing  the sign of the measured hysteresis loops~\cite{PhysRevB.103.035412,PhysRevApplied.20.024065,Li2014Feb,Dankert2018Mar,Tian2015Sep,Ando2014Nov,Yang2016Aug}. But later on it was shown that the sign of the electrically measured spin signal can also depend on the resistance and properties of the electrical contacts~\cite{Li2019May,Vaklinova2018Feb}, making such an assignment ambiguous. 

It would be even more problematic to try to distinguish the two states based on the polarization of the created spin currents. If the TSS exhibited ideal spin-momentum locking, they would yield fully spin-polarized currents (Fig.~\ref{SFigure1}(a)), whereas the polarization caused by RSS is expected to be reduced due to the presence of two Fermi circles with opposite spin orientations (Fig.~\ref{SFigure1}(b))~\cite{Hong2012Aug}. Naively, one therefore might expect stronger spin signals from TSS than from RSS. However, the initial hope of fully spin-polarized currents in TSS got diminished as it was found that strong spin-orbit entanglement reduces the spin polarization to values somewhere above 50\%
\cite{PhysRevLett.105.266806,PhysRevLett.110.216401,ARPES-review,arpes-polarization,PhysRevB.86.235106}.

Furthermore, it is not possible to directly determine the spin polarization of the surface currents by electrical means, as the measured spin signal depends both on this spin polarization but also on the spin detection efficiencies of the electrical contacts~\cite{ISI000249789600001}. On the other hand, it is not possible to independently determine these spin detection efficiencies. Instead, these efficiencies are normally calculated from the measured spin signal under the assumptions of specific spin transport and detection models~\cite{Ahn2020,HAN2012369,PhysRevLett.105.167202,PhysRevB.80.214427}. We are not aware of any kind of estimation of spin detection efficiencies in devices with topological insulators, but in graphene-based devices with similar contact materials these efficiencies can vary quite significantly from almost zero to around 30\%~\cite{Ahn2020,HAN2012369,PhysRevLett.105.167202,PhysRevB.80.214427}.

To make matters even more complicated, TSS and RSS can coexist in the same device (see Fig.~\ref{SFigure1}(c)) both contributing to to the measured spin voltage at the detector. Furthermore, so far it has been implicitly assumed that the same states are responsible for charge transport in both the transport channel between the contacts and the topological insulator directly below the contacts. However, in recent years several studies argued that invasive contacts might disturb the spin polarized surface states and are therefore likely responsible for diminished spin signals ~\cite{Culcer2020,Majumder2017,Walsh2017,PhysRevB.90.085115,Longo2025}.
Additionally, contact-induced charge transfer and band bending in the first few quintuple layers of the TI is argued to create additional RSS underneath the contacts~\cite{PhysRevB.90.085115,Yang2016Aug}. Overall, the combination of varying spin detection efficiencies, unknown spin polarization of TSS, the possible coexistence of both TSS and RSS in the transport channel, and the poorly understood impact of contacts on the surface states make an assignment of the measured spin signal to either TSS or RSS based on the sign and/or the amplitude of the measured spin signal inconclusive.

An approach to circumvent the problem of the varying and unknown impact of contacts is to use electrostatic gating to tune the Fermi-level within one specific device. For example, in Ref.~\cite{PhysRevApplied.20.024065} it was argued that a gate-dependent sign reversal in the measured spin signal can accordingly be attributed to a transition from a combination of RSS- and TSS-dominated transport (black dashed line in Fig.~\ref{SFigure1}(c)) to a solely TSS-dominated transport (gray dashed line in Fig.~\ref{SFigure1}(c)). However, this approach is based on the assumption that the surface states remain unchanged by electrostatic gating. But angle-resolved photoemission spectroscopy measurements demonstrated that the Rashba-induced spin splitting of the two-dimensional electron gas can be influenced by electrostatic control and adsorbates~\cite{PhysRevLett.107.096802,PhysRevB.111.165115} or by
optical excitation of charge carriers~\cite{Michiardi2022}. Overall, it was found that the strength of the spin-orbit coupling and the resulting spin polarization of both RSS and TSS can depend on subtle variations in the spatial charge density~\cite{PhysRevB.93.035110}. Unfortunately, most doping-dependent changes of RSS and TSS were measured by the evaporation of adsorbates onto the surface of the topological insulator and not by electrostatic gating. Therefore, it remains to be seen how great the impact of electrostatic gating on RSS and TSS is and how this might impact the interpretation of gate-dependent transport measurements in one specific device.